\documentclass[twocolumn,showpacs,amsmath,amssymb,showkeys,floatfix,prl]{revtex4}
\usepackage{amsfonts,amsmath}
\usepackage{graphicx}
\usepackage{dcolumn}
\usepackage{bm}

\begin{document}

\title{Three-neutrino model analysis of the world's oscillation data}

\author{D. C. Latimer and D. J. Ernst}

\affiliation{Department of Physics and Astronomy, Vanderbilt University, 
Nashville, Tennessee 37235, USA}
\date{\today}

\begin{abstract}
A model of neutrino oscillation experiments is constructed. The experiments 
incorporated are: solar neutrinos (Chlorine, Gallium, Super-K, and 
SNO), reactor neutrinos (Bugey and CHOOZ), beam stop neutrinos (LSND decay at 
rest and decay in flight), and atmospheric neutrinos. Utilizing this model and 
the standard three-neutrino mixing extension of the standard model, the data are
analyzed. Solutions for the mixing angles and mass-squared differences
are found to occur in 
pairs corresponding to the interchange $\Delta m^2_{12} \leftrightarrow
\Delta m^2_{23}$. Two pairs of solutions are found that reasonably reproduce 
the data, including the LSND data. These solutions are 
$\theta_{12}\approx 0.5$, 
$\theta_{13}\approx 0.1$, $\theta_{23}\approx 0.7$, $\Delta m^2_{12} \approx 
5\times 10^{-5}$ eV$^2$ and $\Delta m^2_{23} \approx 0.2$ eV$^2$ or  
2.4 eV$^2$. Other statistically significant solutions are also found which 
produce negligible oscillations for the LSND experiments.
\end{abstract}



\maketitle

Evidence for neutrino oscillations arises from solar neutrino experiments
\cite{homestake,sage1,sage2,gallex,gno,superksolar1,sno1,sno2}, neutrinos
emitted from reactors \cite{bugey,chooz1,chooz2,kamland1}, 
neutrinos from the beam stop of an accelerator \cite{lsnd1,lsnd2}, 
and neutrinos originating from cosmic rays impinging on the atmosphere 
\cite{sk1,sk2,sk3,sk4}. We here model what we believe to be the essential 
physics of each of these experiments. We analyze this model utilizing the 
standard three neutrino mixing extension of the standard model. We find 
a number of sets of parameters, mixing angles and mass-squared differences, 
which reproduce the data, some of which reproduce the entire data set including 
the LSND experiments. Previous examinations of three neutrino mixing have either 
excluded the LSND experiments \cite{gonzalezfit1, gonzalezfit2}, limited
the mass-squared differences \cite{ohlsson1}, or used approximations and 
constraints in order to work analytically \cite{ahluwalia1,ahluwalia2,
ahluwalia3,weiler}.

Neutrino oscillations arise because neutrinos are created in flavor eigenstates, 
and the flavor eigenstates are not equal to the mass eigenstates. The flavor
eigenstates (labeled by $\alpha$) are related to the mass eigenstates 
(labeled by $k$)
through a unitary matrix $U_{\alpha k}$,

\begin{widetext}
\begin{equation}
U_{\alpha k} \rightarrow \left(  
\begin{array}{ccc}
c_{12} c_{13} & s_{12} c_{13} & s_{13} \\
-s_{12}c_{23} - c_{12} s_{23} s_{13}  & 
c_{12} c_{23} - s_{12} s_{23} s_{13}  & s_{23}c_{13}\\
 s_{12}c_{23} - c_{12} c_{23} s_{13}  & 
-c_{12} s_{23} - s_{12} c_{23} s_{13}  & c_{23}c_{13}
\end{array}
\right)\,\,,
\end{equation}
\end{widetext}
where $c_{\alpha k} = \cos{\theta_{\alpha k}}$,
$s_{\alpha k}=\sin{\theta_{\alpha k}}$, and $\theta_{\alpha k}$ is real.
We assume no CP violation. We order the mass 
eigenstates by decreasing mass, and the flavor eigenstates are 
ordered electron, mu, tau. The probability that a neutrino of flavor $\alpha$ 
will be detected a distance $L$ away as a neutrino of flavor $\beta$
is given by
\begin{eqnarray}
P_{\alpha \to \beta}(L/E) &=& \delta_{\alpha \beta} \nonumber\\  
&&-4\sum_{\genfrac{}{}{0pt}{}{j,k=1}{j< k}}^3
U_{\alpha j} U_{\beta j} U_{\alpha k} U_{\beta k} \sin^2 \phi^{osc}_{jk}\,\,,
\label{pee}
\end{eqnarray}
with $\phi^{osc}_{jk}=1.27 \Delta m_{jk}^2 L/E$, where the units of $L/E$ 
are m/MeV, and $\Delta m_{jk}^2 \equiv
m_j^2 - m_k^2$ in units of eV$^2$. 

For the neutrino flux emitted by the sun we use the standard solar model
\cite{bp2000}. A number of detectors have measured the flux
of solar neutrinos arriving here on Earth. Each has a different acceptance and 
thus measures different energy neutrinos. Each measures a deficit as compared 
to the flux predicted by the standard solar model.
In order to reproduce the energy dependence of the 
survival rate of electron neutrinos arriving at the Earth as 
seen in the experiments, we must invoke the 
MSW effect \cite{{wolf1,ms}}. The MSW effect arises because the neutrinos created in 
the sun propagate through a medium with a significant electron density. The 
forward coherent elastic neutrino-electron scattering produces an effective 
change, relative to the mu and tau neutrino, in the mass of the electron
neutrino given by 
$A(r)=\sqrt{2} \,G \,E \,\rho(r)/m_n$, with $\rho(r)$ the electron density at a 
radius
$r$, $G$ the weak coupling constant, and $m_n$ the nucleon mass. In the flavor basis, the Hamiltonian 
then becomes 
\begin{equation}
H_{mat} = U {\cal M} U^\dagger + A \,\,,
\end{equation}
with ${\cal M}$ the (diagonal) mass-squared matrix in the mass eigenstate basis 
and $A$ the 
$3\times 3$ matrix with the interaction $A(r)$ as the  
electron-electron matrix element and zeroes elsewhere. By diagonalizing this
Hamiltonian with a unitary transformation $D_{\alpha k}(r,E)$, we define local 
masses and eigenstates as a function or $r$ and $E$. Care must be taken so that
$D_{\alpha k}(r,E)$ becomes $U_{\alpha k}$ in the limit of zero electron density.
In the adiabatic limit, which we use, the electron survival probability is 
\begin{equation}
P_{ee}^{ad}(r,E) = \sum_{k=1}^3 D(r,E)_{ek}^2 \,U_{ek}^2\,\,.
\end{equation}

Neutrinos are produced throughout the sun by various reactions, each with its 
own energy spectrum. The surviving neutrinos are then detected by detectors 
which have a different acceptance for each energy of the neutrino. We model this
by taking the survival probability for an electron neutrino in a particular 
experiment to be given by
\begin{equation} 
P_{ee}^{ex}= \sum_{j=1}^N p_j^{ex} \int_0^{R_\odot}f_j(r) \,dr
\int_{E_{thresh}}^\infty g_j(E) \,
P_{ee}^{ad}(r,E)\,dE\,\,.
\end{equation}
Here, $j$ labels a particular nuclear reaction; we include three reactions --
pp, $^7$Be, and $^8$B. The quantity $p_j^{ex}$ is the probability that 
in a particular experiment the neutrino arose from nuclear reaction $j$. We take
these from the analysis of Ref.~\cite{neutreview} for the solar experiments: 
chlorine \cite{homestake},
gallium (GALLEX/GNO,Sage) \cite{sage1,sage2,gallex,gno}, Super-K 
\cite{superksolar1}, and SNO
\cite{sno2}. The function $f_j(r)$ 
is the probability that a neutrino is created by reaction $j$ at a radius $r$ 
\cite{bp2000} of the sun and is integrated from the center of the sun to the 
solar radius $R_\odot$. The function
$g_j(E)$ is the energy distribution of the neutrinos emitted in reaction $j$. 
For $^7$Be this is a delta function at 0.88 MeV. The lower emission line does 
not contribute significantly. For the pp neutrinos, the energy distribution 
times the detector acceptance is a relatively narrow function of energy;
we set $E$ to its average. For $^8$B neutrinos, we 
use the energy distribution from the standard solar model \cite{bp2000}
and numerically perform the integration. 


We check our treatment of solar neutrinos by performing a two neutrino analysis
of only the solar neutrino experiments. We find a minimum for $\tan^2\theta=$ 
0.43 with bounds $0.28\le \tan^2\theta\le 0.68$ and a minimum for $\Delta m^2 =$ 
$4.0\times 10^{-5}$ eV$^2$ with bounds 
$1.0 \times 10^{-5}~{\rm eV}^2 \le \Delta m^2 \le 8.6 \times 10^{-5}~{\rm eV}^2$.
The analysis from Ref.~\cite{bach} gives $\tan^2\theta=$ 
0.44 with bounds $0.36\le \tan^2\theta\le 0.58$ and a minimum for $\Delta m^2 =$ 
$7.0\times 10^{-5}$ eV$^2$ with bounds 
$5.9 \times 10^{-5}~{\rm eV}^2 \le \Delta m^2 \le 9.7 \times 10^{-5}~{\rm eV}^2$.
This is satisfactory for our goal of locating possible solutions that are 
semi-quantitatively  correct. Our errors are necessarily larger than in
a thorough and model-independent analysis as we do not include all of the 
data, such as the measured neutrino energy spectra.
The experimental data for the solar experiments which we fit are given in 
Table~\ref{t2}. We take values from \cite{neutreview} which differ slightly 
from the original analysis.

The reactor experiments that we include are Bugey \cite{bugey}, CHOOZ 
\cite{chooz1,chooz2} and KamLAND \cite{kamland1}. As there is an energy distribution 
for the neutrinos emitted from a reactor, the electron survival probability 
given by Eq.~\ref{pee} must be integrated over this spectrum. For small values 
of $L/E$, the coherent limit, Eq.~\ref{pee} remains correct. For 
sufficiently large values of $L/E$, the incoherent limit, the 
$\sin^2 \phi^{osc}$ term 
averages to 1/2. The transition between these regions 
depends on the details of the energy distribution of the source neutrinos. 
We simplify 
this by using an average value for $L/E$ and by using $\sin^2 \phi^{osc}$ 
for 
$\phi^{osc} < \pi /4$ and setting 
$\sin^2 \phi^{osc} =$1/2 for $\phi^{osc} > \pi /4$. KamLAND is unique among 
these as it 
measures ${\cal P}_{ee}$ where the others set limits. 

\begin{table}
\begin{center}
\begin{tabular}{|l|c|c|c|}  
\hline\hline
~Experiment~ & Measured & $L/E$ (m/MeV) ) & Data \\ \hline\hline
~LSND-DAR & ${\cal P}_{\overline e \overline\mu}$ & .73 &$3.1 \pm 1.3 \times 10^{-3}$ \\ \hline
~LSND-DIF & ${\cal P}_{e\mu}$ & .40 &$2.6 \pm 1.1 \times 10^{-3}$ \\ \hline
~CHOOZ   & ${\cal P}_{ee} $ & 300. &$> 0.96$ \\ \hline
~Bugey   & ${\cal P}_{ee} $ & 10.3 &$> 0.95$ \\ \hline
~KamLAND & ${\cal P}_{ee} $ & $4.1 \times 10^4$ &$.611 \pm .094 $ \\ \hline
~Super-K & ${\cal P}_{ee} $ & $2.2 \times 10^{10}$ &$.465 \pm .094 $ \\ \hline
~SNO     & ${\cal P}_{ee} $ & $2.2 \times 10^{10}$ &$.348 \pm .073 $ \\ \hline
~Chlorine& ${\cal P}_{ee} $ & $4.0 \times 10^{10}$ &$.337 \pm .065 $ \\ \hline
~Gallium & ${\cal P}_{ee} $ & $35. \times 10^{10}$ &$.550 \pm .048 $ \\ \hline
~Atmospheric    & ${\cal R}_e^{^{E<1}} $ & $1.9 \times 10^4$ &$1.13 \pm .09 $ \\ 
        &                        & 21.4 &                \\ \hline
~Atmospheric    & ${\cal R}_e^{^{E>1}} $ & $2.6 \times 10^3 $&$.85 \pm .16 $ \\ 
        &                        & 3.0               &   \\ \hline
~Atmospheric    & ${\cal R}_\mu^{^{E<1}} $ &$1.6 \times 10^4$ & $.73 \pm .06 $ \\ 
        &                          &18.8             &    \\ \hline
~Atmospheric    & ${\cal R}_\mu^{^{E>1}} $ &$4.3 \times 10^4$& $.62 \pm .09 $ \\ 
        &                          &5.0              &    \\ \hline
\end{tabular}
\end{center}
\caption{The experiments, quantity measured, the average value of $L/E$, and 
experimental data for those quantities included in the model are presented.
For the atmospheric 
data, the quantity ${\cal R}$ is define in the text, and the upper (lower) 
value of $L/E$ is for upward  (downward) going neutrinos. Note that Super-K 
plus SNO combined provide a measurement of ${\cal P}_{ee}$ and ${\cal P}_{ee}+
{\cal P}_{e \mu}+{\cal P}_{e \tau}$.}
\label{t2}
\end{table}

The LSND experiments 
use neutrinos produced from muons created in the LAMPF beam stop. There are 
two experiments. The decay at rest experiment \cite{lsnd1}
measures the oscillation of an 
muon antineutrino into an electron antineutrino, while the decay in flight
experiment \cite{lsnd2} 
measures the oscillation of a muon neutrino into an electron
neutrino. These experiments are unique in that they measure the appearance
of a different flavor neutrino rather than the disappearance of electron 
neutrinos. We treat the $\sin^2 \phi^{osc}$ in Eq.~\ref{pee} just as we did for the 
reactor neutrinos.

The Super-Kamiokande experiment \cite{sk1,sk2,sk3,sk4} has also measured neutrinos that
originate from cosmic rays hitting the upper atmosphere. The detector 
distinguishes between $e$-like (electron and anti-electron) neutrinos and 
$\mu$-like (muon and anti-muon) neutrinos.The rate of $e$-like
neutrinos of energy $E$ arriving at the detector from a source a distance $L$ 
away is
\begin{equation}
R_e (L,E) = {\cal P}_{ee}(L,E) + n(E) {\cal P}_{e\mu}(L,E)\,\,,
\end{equation}
and for $\mu$-like neutrinos
\begin{equation}
R_\mu (L,E) = {\cal P}_{\mu \mu}(L,E) + \frac{1}{n(E)} {\cal P}_{e\mu}(L,E)\,\,,
\end{equation}
where $n(E)$ is the ratio of $\mu$-like neutrinos to $e$-like neutrinos at the 
source. We separate the
data into neutrinos with an energy less than or greater than 1 GeV. An average 
value for the energy is calculated from results given in Ref.~\cite{sk2,sk3} 
which uses the model of the neutrino fluxes from Ref.~\cite{hondaflux}. 
We approximate the energy distribution of these neutrinos by fitting a simple 
``teepee'' shaped function to the distributions given in Ref.~\cite{sk2,sk3}. 
This allows us to do the energy integral analytically. The energy averaged 
values of $R_\alpha$ we call $R_\alpha (r,\langle E \rangle )$ The high-energy 
$\mu$-like events are classified as ``fully contained'' or ``partially 
contained'' events and each of these arises from 
a different energy spectrum. We combined these as 0.24 fully contained and
0.76 partially contained \cite{sk4}. 

To remove the dependence on the overall normalization of 
neutrino flux, the ratio of measured fluxes for upward going (coming from the
far side of the earth) neutrinos to downward going (coming from overhead) 
neutrinos is taken. This ratio is
\begin{equation}
{\cal R}_\alpha = \frac{R_\alpha (r_{up},\langle E \rangle)}
{R_\alpha (r_{down},\langle E \rangle)}\,\,.
\end{equation}
We utilize a definition of upward/downward going
neutrinos as those with the scattered lepton direction in the detector
of no more than $\pi/5$ radians off-axis. 
The downward going neutrinos were 
assumed to travel 15 km from the top of the atmosphere, 
whereas the upward ones travel one earth diameter, 13,000 km.  
The experiment measures the neutrino fluxes as a function of 
the azimuthal angle. We utilize only the endpoints of these 
measurements. Assuming that if we fit the endpoints, the curve between would 
equally well be fit, we divide the error associated with
the endpoint by $\sqrt {5/2}$ (5/2 $=$ the number of experimental points over 
the number used) to more properly weight this experiment with respect to the 
others. The data and the parameters used for all the experiments are given in 
Table~\ref{t2}.

We fit the mixing angles and the mass-squared differences to the data by 
minimizing chi-squared per degree of freedom $\chi^2_{dof}$. In 
Table~\ref{t3}, we present the parameters for the ten best fits. We have 
also found local minima for $\chi^2_{dof}$ near 2.3 and 2.7. Notice that
the solutions come in pairs, (solutions 1 and 4, 2 and 3, 5 and 6, 7 and 8, 
9 and 10)
corresponding to the interchange 
$\Delta m^2_{12} \leftrightarrow \Delta m^2_{23}$ and an appropriate 
redefinition of the mixing angles. We derive this symmetry elsewhere 
\cite{sym}.

The experimental data and the theoretical results for these fits are presented
in Table \ref{t4}. Note that solutions 1 through 4 produce non-negligible and 
reasonable results for the LSND experiments while fitting the remainder of the 
experiments. We can see how this comes about by first considering the solar 
experiments. We find a mass-squared difference on the order of $10^{-5}$ eV$^2$,
which is the magnitude needed for the MSW effect to produce the energy
dependence of the solar neutrinos. Quantitatively, this is different from the 
$2 \times 2$ case as the MSW effect is producing an energy dependence for a case 
where ${\cal P}_{ee}(\infty )$ is about 0.6. In fact, we find that KamLAND is
measuring ${\cal P}_{ee}(\infty )$.

\begin{table}
\begin{center}
\begin{tabular}{|c|c|c|c|c|c|c|} \hline\hline
~fit~~& ~$\chi^2_{dof}$~ & ~$\theta_{12}$~ & ~$\theta_{13}$~ & ~$\theta_{23}$~ &
 $\Delta m^2_{12}$ (eV)$^2$ & $\Delta m^2_{23}$ (eV)$^2$ \\ 
\hline\hline
1 & 1.4 & .55 & .16 & .79 &$5.6 \times 10^{-5}$ & .24 \\ \hline
2 & 1.5 & .58 & .06 & .66 & $4.3\times 19^{-5}$ & 2.3 \\ \hline
3 & 1.6 & 1.5 & .59 & .99 & 2.3 & $4.6 \times 10^{-5}$ \\ \hline
4 & 1.7 & 1.4 & .58 & .99 & .24 & $5.0 \times 10^{-5}$ \\ \hline
5 & 1.9 & .58 & .10 & .78 & $4.2 \times 10^{-5}$ & .12 \\ \hline
6 & 1.9 & .58 & .10 & .92 & $.12 $ & $4.0\times 10^{-5}$ \\ \hline
7 & 2.0 & .63 & .001& .37 & $1.4 \times 10^{-5}$&$1.2 \times 10^{-3}$ \\ \hline
8 & 2.0 & $\pi /2$ & .63 & 1.2 & $1.2\times 10^{-3}$ & $1.4 \times 10^{-5}$ 
  \\ \hline
9 & 2.0 & .63 & .05 & .63 & $1.4 \times 10^{-5}$& .11  \\ \hline
10& 2.0 & $\pi /2$ & .63 & .96 & .11 & $1.4 \times 10^{-5}$ \\ \hline
\end{tabular}
\end{center}
\caption{The value of $\chi^2_{dof}$ and the fit parameters for the 
ten best fits.}
\label{t3}
\end{table}
 
Considering LSND and the reactor experiments Bugey \cite{bugey} and CHOOZ
\cite{chooz2}, we note that the term in 
Eq.~\ref{pee} with $\Delta m^2 \approx 10^{-5}$ eV$^2$ does not contribute. The 
coefficients of $\sin^2 \phi^{osc}$ for ${\cal P}_{e\mu}$ for the 
large mass-squared difference terms are each of the
order $10^{-2}$. They are of opposite sign and equal to each other to about 
ten percent. This produces the LSND results for ${\cal P}_{e\mu}$ of order 
$10^{-3}$. For
${\cal P}_{ee}$ these coefficients are of the same sign but are individually 
of the order of $10^{-3}$ and thus do not give results which contradict Bugey
or CHOOZ. The coefficient of the $\sin^2 \phi^{osc}$ term corresponding to the 
mass-squared difference of $10^{-5}$ eV$^2$ term is about 0.25. This term then 
contributes significantly to the asymptotic value of ${\cal P}_{ee}$ which is
important for fitting KamLAND and the solar experiments. For solutions 1 and 4,
$\phi^{osc}$ for the large mass-squared difference terms for LSND experiments 
are in the coherent region, while for solutions 2 and 3 they are near $\pi /2$.

Finally we look at the atmospheric data. We find that for solutions 2 and 3, 
the solutions with the largest mass-squared difference, all the values of 
$\phi^{osc}$ are greater than $\pi/ 2$. This does not give a perfect fit to the
atmospheric data, but it is sufficiently close that when combined with an 
excellent fit to the LSND data, a good $\chi^2_{dof}$ results. For solutions
1 and 4, the large mass-squared difference is smaller in order to better fit the 
atmospheric data, but at a cost to the fit of LSND. 
 
We also note that our solution 7 corresponds to that found in 
Ref.~\cite{gonzalezfit2}, 
in which LSND was not included. This helps give us confidence that our 
simplified model of the experiments is capable of locating possible solutions.

We have found a set of mixing angles and mass-squared differences that, within a
model, can produce results that fit the world's
data, including the LSND experiments. The future requires a thorough and 
model-independent analysis to see whether this is actually so. We have also 
found a
new symmetry that arises from the interchange of the mass-squared 
differences.

\begin{acknowledgments}
The authors are grateful for very helpful conversations with D. V. Ahluwalia and
I. Stancu. This work is supported by the U.S. Department of Energy under 
grant No.~DE-FG02-96ER40963.
\end{acknowledgments}

\bibliographystyle{unsrt}
\bibliography{neutfitf}

\begin{widetext}
\begin{center}
\begin{table}
\begin{center}
\begin{tabular}{|c|c|c|c|c|c|c|c|c|c|c|c|} \hline\hline
Experiment& Data & 1 & 2 & 3 & 4 & 5 & 6 & 7 & 8 & 9 & 10 \\ \hline\hline
LSND-DAR ($\times 10^{-3}$)& $3.1\pm 1.3$ & 2.4 & 2.8 & 2.7 &
    1.5 & .23 & .13 & .00 & .00 & .04 & .04 \\ \hline
LSND-DIF ($\times 10^{-3}$)& $2.6\pm 1.1$ & .74 & 2.8 & 2.7 &
    .45 & .07 & .04 & .00 & .00 & .01 & .00 \\ \hline
CHOOZ & $>.96$ & .95 & .99 & .99 & .96 & .98 & .99 & 1.0 & 1.0 & .99 & 1.0
     \\ \hline
Bugey & $>.95$ & .95 & .99 & .99 & .96 & .98 & .99 & 1.0 & 1.0 & .99 & 1.0
     \\ \hline
KamLAND & $.611\pm .094$ & .58 & .57 & .57 & .55 & .57 & .57 & .60 & .60 & .59
     & .59 \\ \hline
Super-K & $.465\pm .094$ & .33 & .34 & .35 & .34 & .33 & .34 & .35 & .35 & .35
     & .35 \\ \hline
SNO     & $.348\pm .075$ & .33 & .34 & .35 & .34 & .33 & .34 & .35 & .35 & .35
     & .35 \\ \hline
Chlorine& $.337\pm .065$ & .39 & .39 & .40 & .39 & .38 & .38 & .37 & .37 & .37
     & .37 \\ \hline
Gallium & $.550\pm .048$ & .54 & .54 & .54 & .52 & .53 & .53 & .49 & .49 & .49
     & .49 \\ \hline
${\cal R}_e^{^{E<1}} $ &$1.13\pm .09$& 1.17 & 1.27 & 1.26 & 1.15 & 1.12 & 
      1.12 & 1.12 & 1.12 & 1.06 & 1.06 \\ \hline
${\cal R}_e^{^{E>1}} $ &$.85\pm .16 $& 1.04 & 1.06 & 1.07 & 1.05 & 1.05 & 
      1.05 & 1.01 & 1.01 & 1.01 & 1.01 \\ \hline
${\cal R}_\mu^{^{E<1}}$& $.73\pm.06$ & .79  & .85  & .86  & .82  & .73  & 
       .73 &  .73 &  .73 & .73  &  .72 \\ \hline
${\cal R}_\mu^{^{E>1}}$ &$.62\pm .09$& .64  & .69  & .70  & .66  & .62  & 
      .63  &  .62 &  .62 & .65  &  .66 \\ \hline
\end{tabular}
\end{center}
\caption{The experimental results and the predictions of the model for the ten
fits given in Table \protect\ref{t3}.}
\label{t4}
\end{table}
\end{center}
\end{widetext}
\end{document}